\documentclass[3p,times,twocolumn]{elsarticle}

\usepackage{ecrc}

\volume{00}

\firstpage{1}

\journalname{Nuclear Physics B Proceedings Supplement}

\runauth{D. Caprioli}

\jid{nuphbp}

\jnltitlelogo{Nuclear Physics B Proceedings Supplement}

\usepackage{amssymb}

\usepackage{caption}

\usepackage[figuresright]{rotating}
\renewcommand{\deg}{^{\circ}}

\begin{document}

\begin{frontmatter}



\dochead{}

\title{Hybrid Simulations of Particle Acceleration at Shocks}


\author{Damiano Caprioli}

\address{Princeton University -- 4 Ivy Ln., 08544 Princeton NJ, USA}

\begin{abstract}
We present the results of large hybrid (kinetic ions - fluid electrons) simulations of particle acceleration at non-relativistic collisionless shocks.
Ion acceleration efficiency and magnetic field amplification are investigated in detail as a function of shock inclination and strength, and compared with predictions of diffusive shock acceleration theory, for shocks with Mach number up to 100.
Moreover, we discuss the relative importance of resonant and Bell's instability in the shock precursor, and show that diffusion in the self-generated turbulence can be effectively parametrized as Bohm diffusion in the amplified magnetic field.
\end{abstract}


\end{frontmatter}


\section{Introduction\label{sec:intro}}
Astrophysical collisionless shocks are usually associated with non-thermal emission, efficient particle acceleration, and magnetic field enhancement. 
The most prominent examples of non-relativistic collisionless shocks are the blast waves of supernova remnants (SNRs), which are thought to be the sources of Galactic cosmic rays (CRs) up to $\sim10^{17}$eV.
Particles are energized by repeatedly scattering across the shock, in a process called  \emph{diffusive shock acceleration} \cite[DSA, e.g.,][]{bell78a,blandford-ostriker78}.
The current carried by energetic ions propagating into the upstream excites plasma instabilities, which lead to the to the generation of magnetic turbulence.
Such amplified magnetic fields enhance the ion scattering, allowing CRs to rapidly gain energy.  

The intrinsic non-linearity of this interplay between energetic particles and the electromagnetic fields in the regime of strong amplification cannot be described with analytical techniques, and numerical ones are needed.
First-principles kinetic simulations (as particle-in-cell, PIC, simulations) follow both electrons and ions, but are computationally very challenging for realistic mass ratios; they allow the simulation of rather limited physical time and length scales, in units of ion gyration and plasma scales.
To overcome this limitation, it is possible to exploit a \emph{hybrid} technique, which models electrons (assumed massless) as a neutralizing fluid, focusing all the computational dynamical range only on the ion dynamics \cite[see][for a review]{Lipatov02}. 

In this work, we summarize the main results of recent, state of the art, hybrid simulations with unprecedentedly-large boxes, exploring the space of environmental parameters relevant for SNR blast waves.
The crucial questions we address are: 
i) the efficiency of DSA, and its dependence on shock strength and geometry;
ii) the effectiveness of magnetic field amplification in the shock precursor, and the nature of the excited turbulence;
iii) the enhancement of particle scattering due to the self-generated turbulence.
These three main topics correspond to three papers by Caprioli \& Spitkovsky \cite{DSA,MFA,diffusion}, which form a cycle of works aimed to systematically study several aspects of particle acceleration at non-relativistic shocks.

\section{\label{sec:acc} Acceleration Efficiency}
All the simulations are performed with the Newtonian \emph{dHybrid} code \citep{gargate+07}, and the shock is setup as outlined in \cite{DSA}.
Lengths are measured in units of $c/\omega_p$, where $\omega_p=\sqrt{4\pi n e^2/m}$ is the ion plasma frequency, and time in units of inverse cyclotron frequency $\omega_c^{-1}=mc/eB_0$, with $c$ the speed of light, $B_0$ the initial magnetic field, and $n, e, m$ the ion density, charge, mass;
velocities are normalized to the Alfv\'en speed $v_A=B_0/\sqrt{4\pi m n}$, and energies to $E_{sh}\equiv m v_{sh}^2/2$, where $v_{sh}$ is the velocity of the upstream fluid in the downstream frame.
The shock strength is expressed by the Alfv\'enic Mach number $M_A\equiv v_{sh}/v_A$.
We assume the sound speed to be comparable to $v_A$, and throughout the paper we indicate both the Alfv\'enic and the sonic Mach numbers simply with $M$.
The shock inclination is defined by the angle $\vartheta$ between the shock normal and the background magnetic field ${\vec B}_0$, so that $\vartheta=0\deg$ for a parallel shock.

\begin{figure}
\centering\includegraphics[trim=0px 0px 0px 0px, clip=true, width=0.5\textwidth]{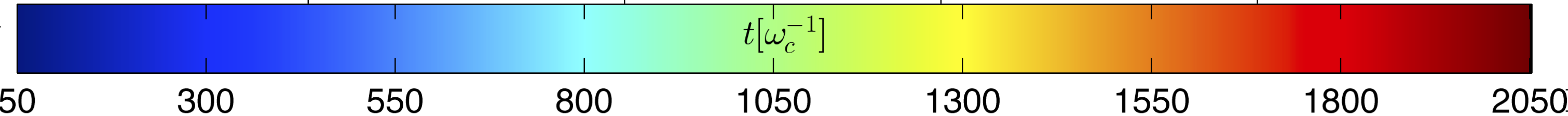}
\includegraphics[trim=0px 15px 0px 270px, clip=true, width=0.5\textwidth]{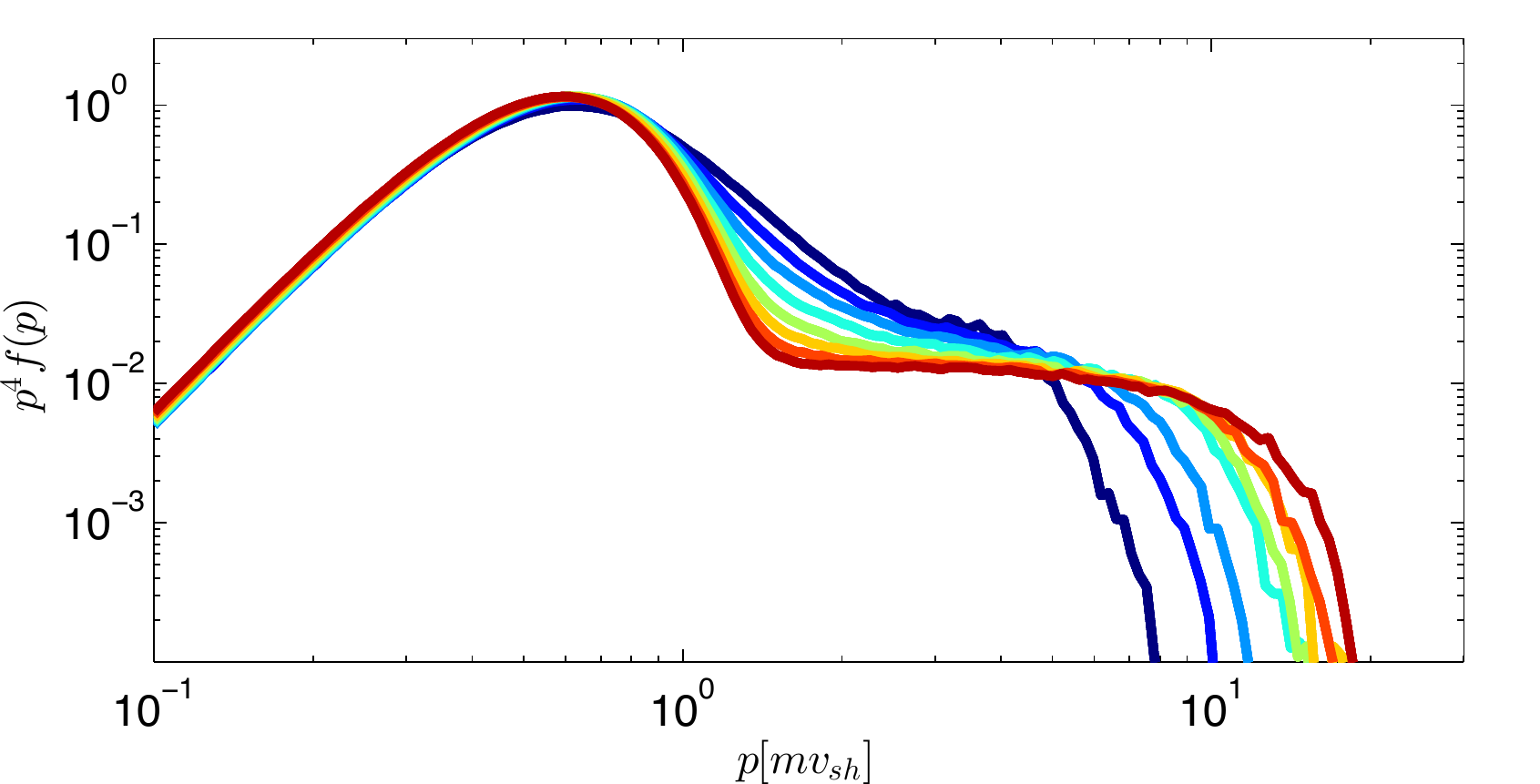}
\caption{\label{fig:p4fp}\footnotesize
Time evolution of the post-shock ion momentum spectrum for a $M=20$ parallel shock, averaged over the whole downstream region.
Notice the peak of the thermal (Maxwellian) distribution for $E\lesssim 2E_{sh}$, and the non-thermal distribution for $E\gtrsim 2E_{sh}$. The spectrum is multiplied by $p^4$ to emphasize the scaling of the power-law tail, which in perfect agreement with DSA prediction \cite{DSA}.}
\end{figure}

As discussed in \citep{DSA}, for $p\gtrsim mv_{sh}$ the ion spectrum develops a non-thermal tail, whose extent (corresponding to the maximum energy achieved by accelerated ions) increases with time (see Figure \ref{fig:p4fp}). 
DSA predicts the spectral slope to depend only on the shock compression ratio $r$ \cite{bell78a,blandford-ostriker78}; since $r\simeq 4$ for $M\gg1$, strong shocks are expected to show universal spectra $\propto p^{-4}$.
The spectrum of non-thermal ions in  Figure \ref{fig:p4fp} agrees perfectly with such a prediction.
More details, and in particular a discussion of the transition between thermal and non-thermal particles can be found in \cite{DSA}.

\begin{figure}\centering
\includegraphics[trim=100px 0px 70px 0px, clip=true, width=0.5\textwidth]{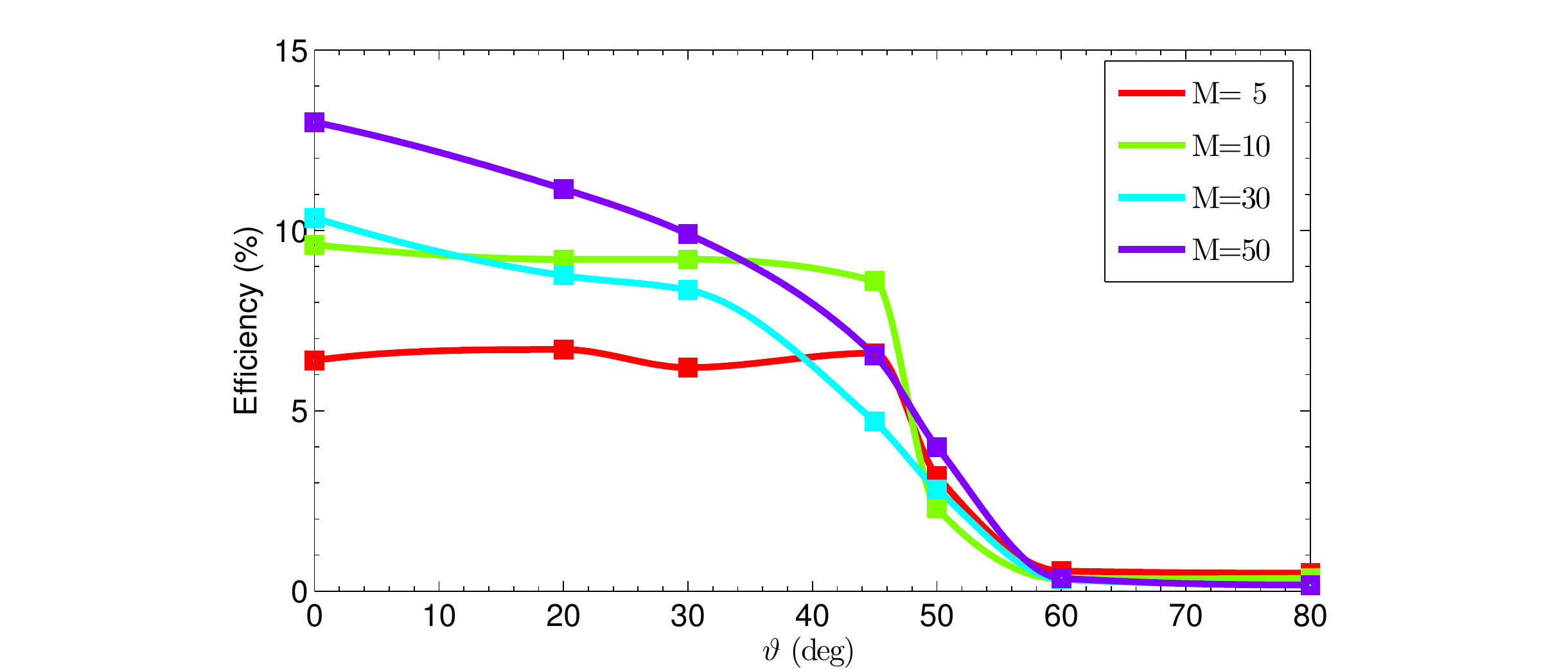}
\caption{\label{fig:eff}\footnotesize
Fraction of the downstream energy density in non-thermal particles at $t=200\omega_c^{-1}$, which represents a good proxy of the saturation value, as a function of shock inclinations and Mach numbers \cite[][]{DSA}.
The largest acceleration efficiency is achieved for strong, parallel shocks, and drops for $\vartheta\gtrsim 45^{\circ}$ regardless of the Mach number.
}
\end{figure}

Figure \ref{fig:eff} shows the acceleration efficiency, expressed as the fraction of the bulk energy flux converted into particles with energy larger than $\sim 10E_{sh}$, for shocks with different strengths and inclinations. 
We outline two important points: i) the acceleration efficiency is $\gtrsim 10\%$ at strong, quasi parallel shocks. 
In these cases, the post-shock temperature is reduced with respect to the one derived from the standard Rankine--Hugoniot conditions, the thermal energy being necessarily reduced to grant energy conservation;
ii) the acceleration efficiency drops for $\vartheta\gtrsim 45\deg$, independently of the shock Mach number. At oblique shocks particles are accelerated by a factor of a few in energy because of shock drift acceleration, but they are advected downstream, and eventually thermalized, before being able to enter DSA.

We have shown, for the first time in PIC/hybrid kinetic simulations of strong non-relativistic shocks, that DSA at quasi-parallel shocks produces the expected spectrum of non-thermal ions, typically with an efficiency larger than 10\%. 
Moreover, we proved that injection into DSA is suppressed if the shock is very oblique.
These findings, also confirmed in 3D setups, are obtained by using very large computational boxes, in both longitudinal and transverse dimensions, and by choosing very small time steps.
In this context, ``large'' and ``small'' refer to the dynamics of highest-energy ions in the simulation, whose diffusion length must be encompassed, and whose Larmor gyration must be time-resolved \cite[see][for a comparison with the previous literature about hybrid simulations]{DSA}. 

\section{\label{sec:MFA} Magnetic Field Amplification}
Since the initial formulation of the DSA theory \cite[e.g.,][]{bell78a,blandford-ostriker78}, particle acceleration has been predicted to be associated with plasma instabilities, and in particular with the generation of magnetic turbulence at scales comparable with the gyroradii of the accelerated particles (resonant streaming instability). 
More recently, it has been pointed out that some non-resonant, short-wavelength modes may grow faster than resonant ones \citep[non-resonant hybrid, NRH, instability: see][]{bell04}. 
On top of these instabilities, which excite modes parallel to the background magnetic field, a transverse, filamentary mode is expected to grow \cite{rb13,filam}. 
In our hybrid simulations, we attest to the presence of all of the instabilities predicted in the quasi-linear theory, and investigate their evolution into the  non-linear regime, where $\delta B/B_0\gtrsim 1$, and the excited turbulence strongly affects the driving CR current. 
Moreover, we account for the large-scale shock structure, which includes advection, and time- and space-dependent particle distributions. 
These \emph{global} simulations overcome the intrinsic limitations of those in periodic boxes, where currents must be prescribed by hand.

\subsection{Filamentation Instability}
\begin{figure}
\begin{center}$
\begin{array}{l}
\includegraphics[trim=45px 75px 40px 55px, clip=true, width=0.5\textwidth]{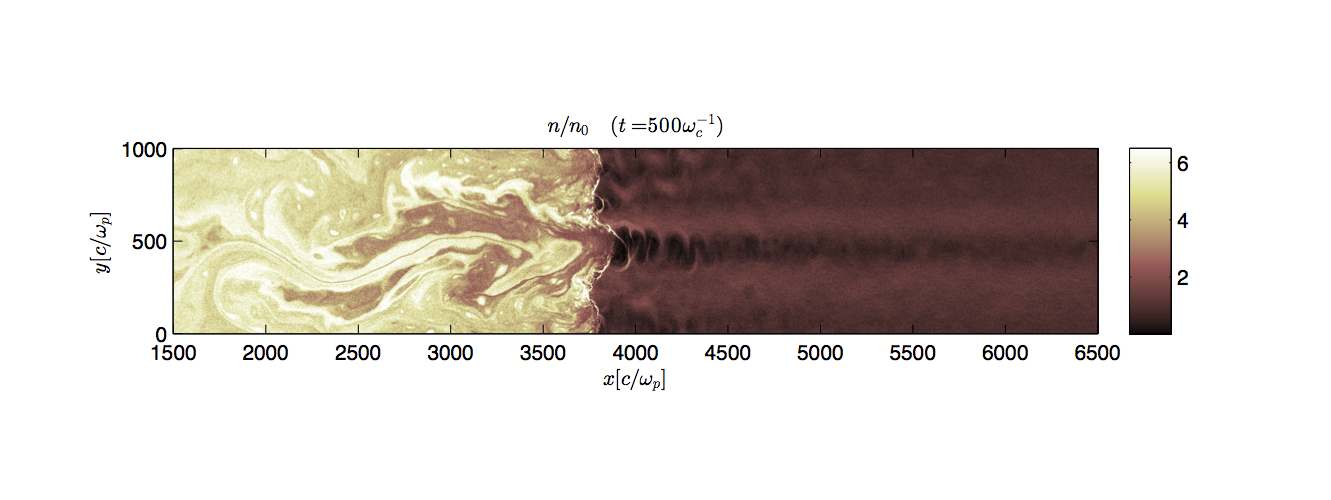} \\
\includegraphics[trim=40px 75px 40px 55px, clip=true, width=0.5\textwidth]{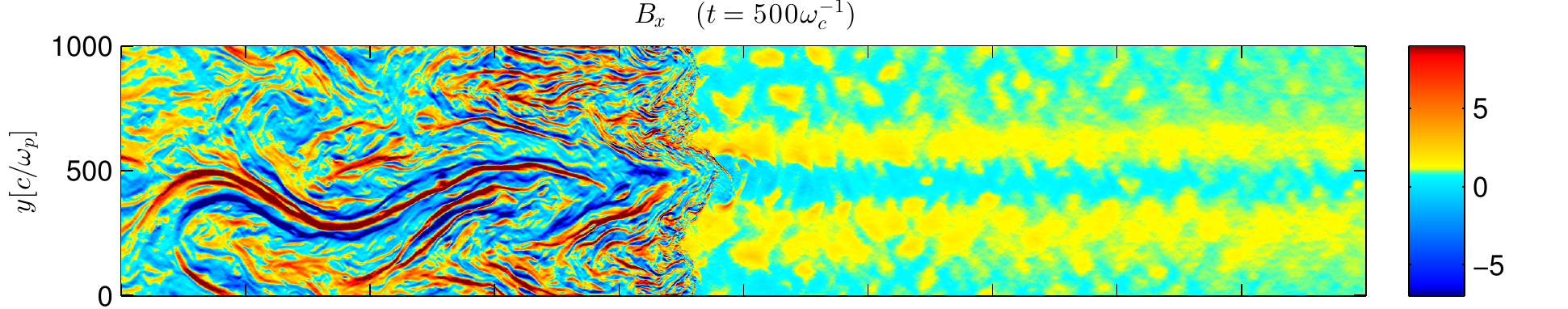} \\
\includegraphics[trim=40px 75px 40px 55px, clip=true, width=0.5\textwidth]{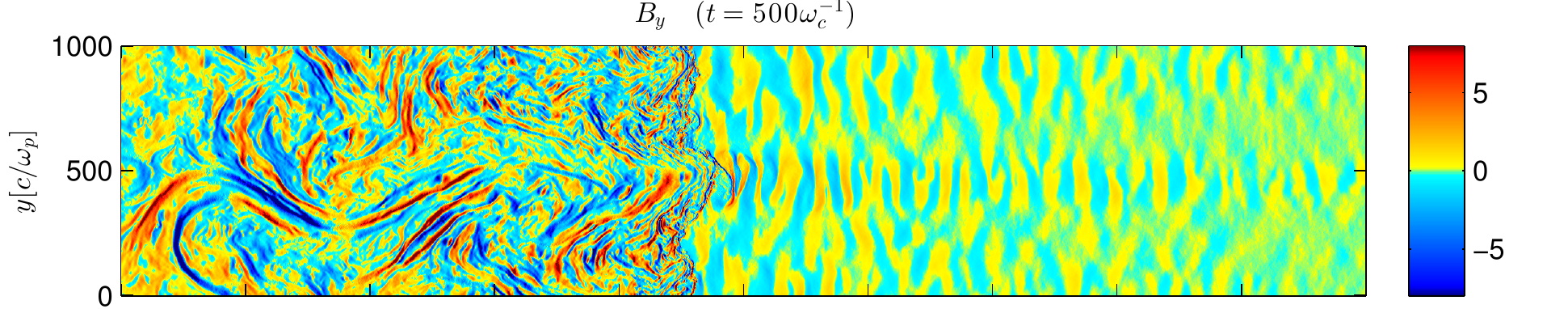} \\
\includegraphics[trim=40px 75px 40px 55px, clip=true, width=0.5\textwidth]{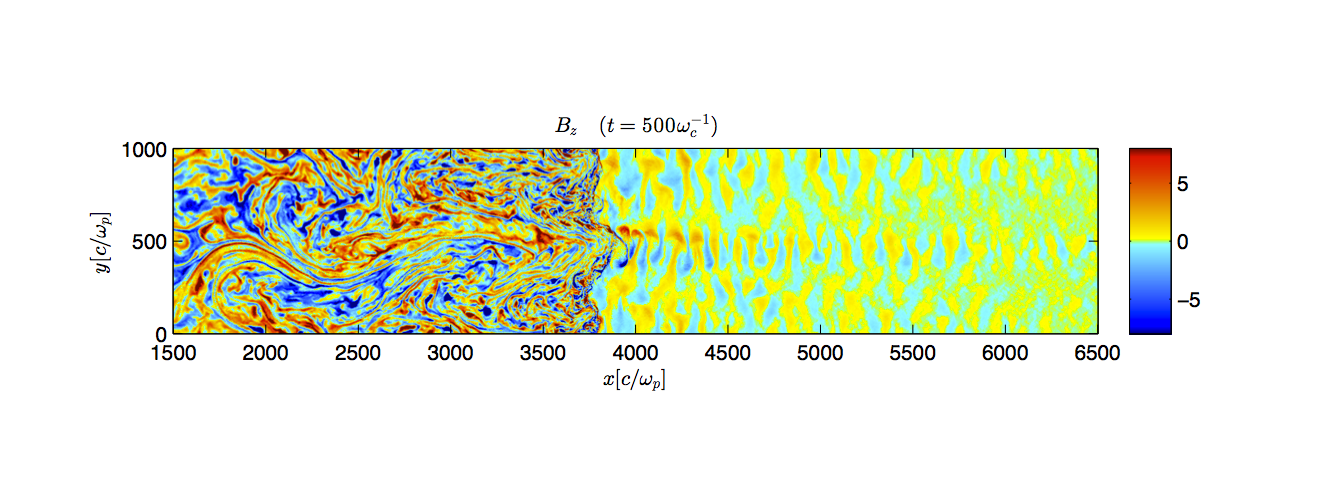} \\
\includegraphics[trim=40px 75px 40px 55px, clip=true, width=0.5\textwidth]{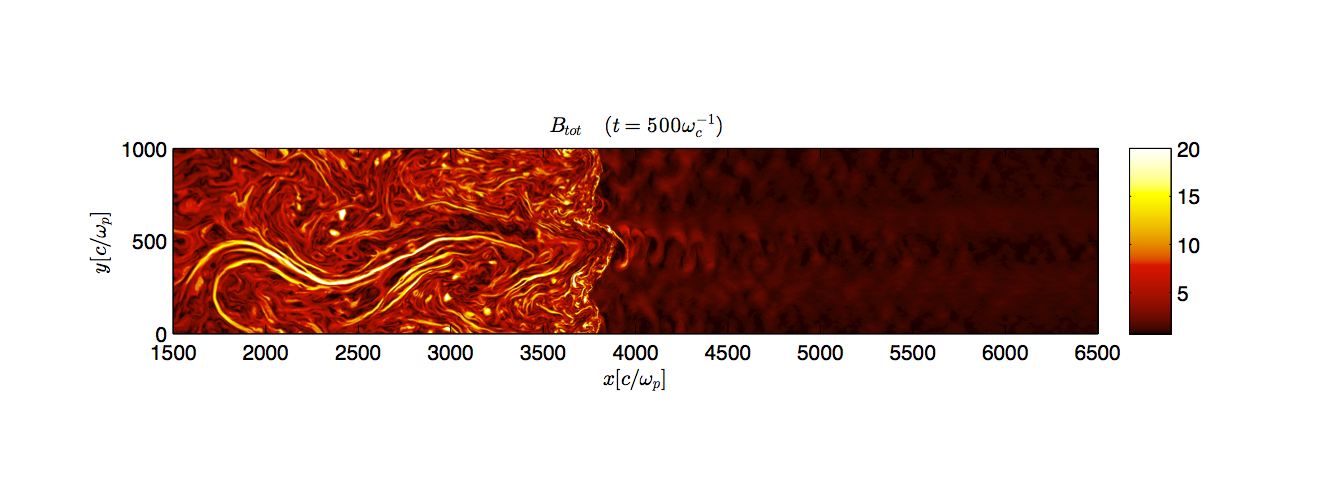}\\
\includegraphics[trim=35px 57px 40px 45px, clip=true, width=0.45\textwidth]{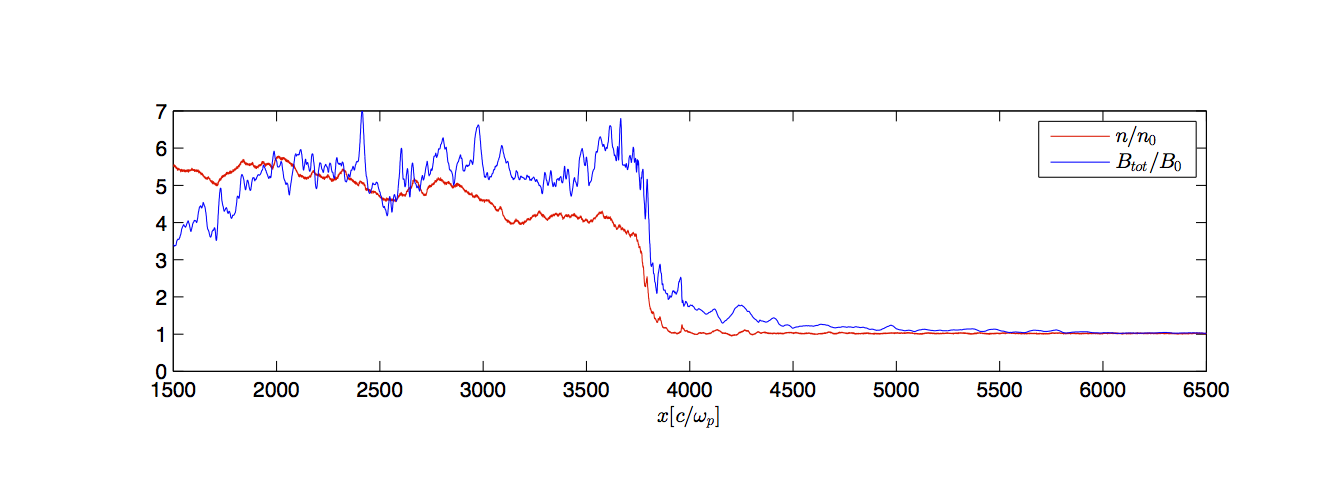} \\
\includegraphics[trim=45px 40px 40px 45px, clip=true, width=0.5\textwidth]{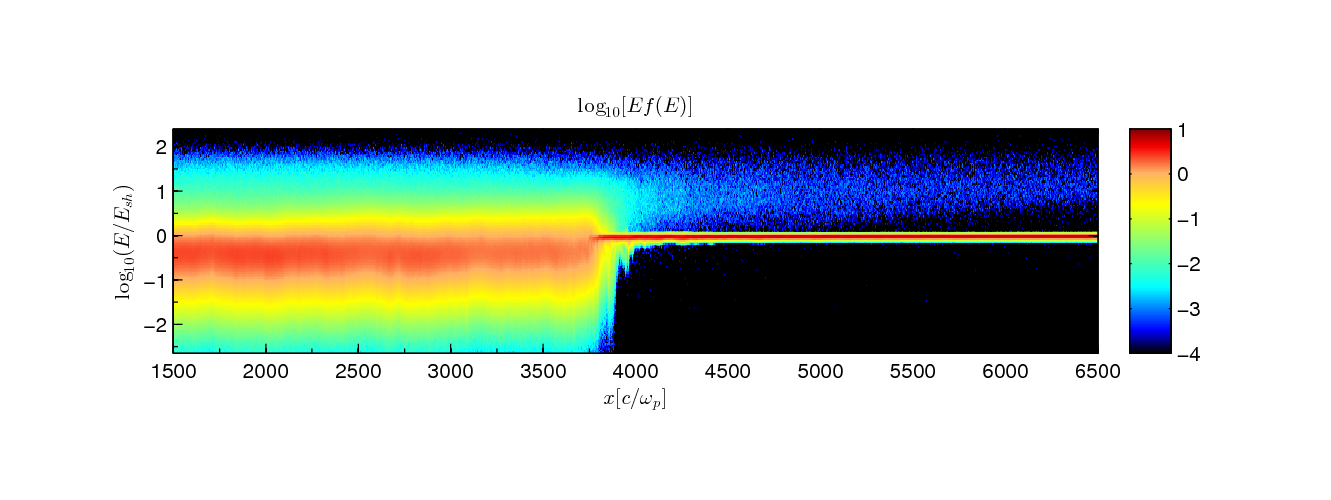} \\
\includegraphics[trim=40px 75px 40px 55px, clip=true, width=0.5\textwidth]{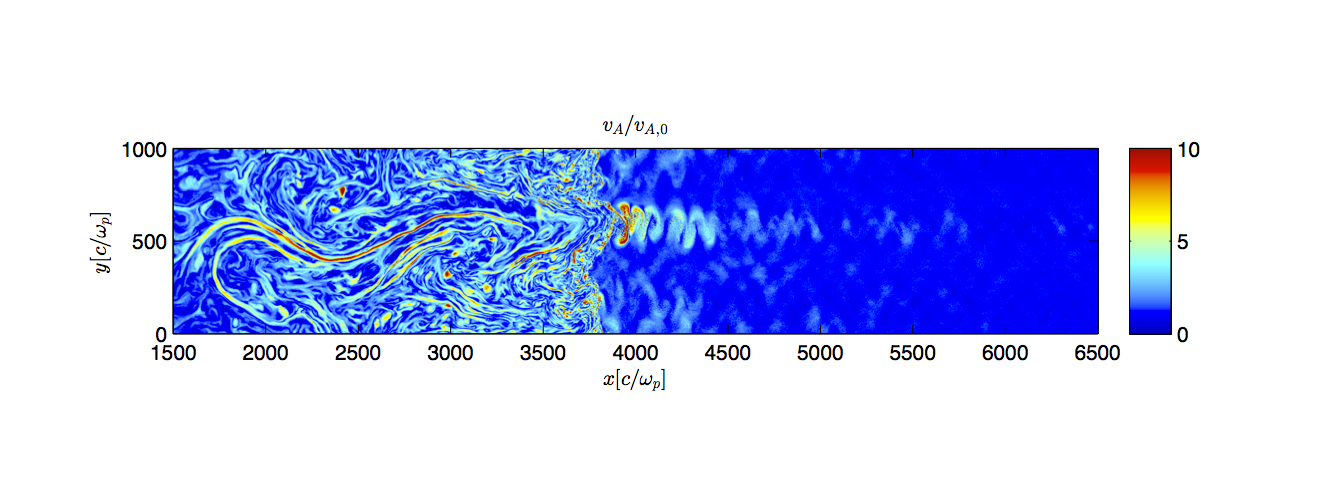} \\
\includegraphics[trim=40px 40px 40px 55px, clip=true, width=0.5\textwidth]{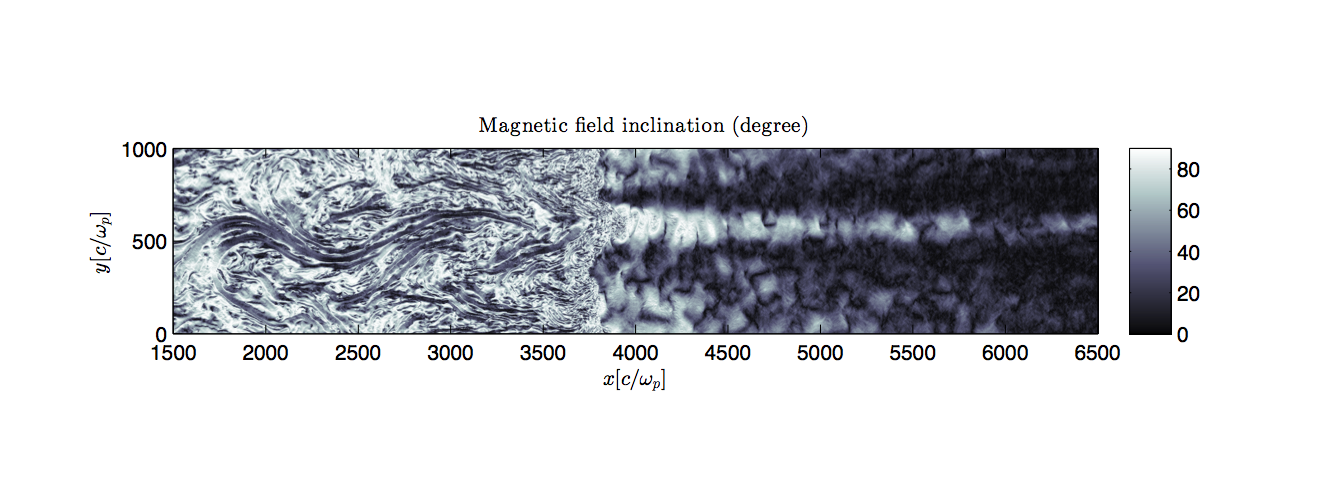} \\
\end{array}$
\end{center}
\caption{\label{fig:hybrid}\footnotesize
Output for a 2D simulation of a parallel shock with $M=30$, at $t=500\omega_c^{-1}$. 
Quantities are plotted as a function of position, and are (from top to bottom): ion density ($n$); three components of the magnetic field ($B_x$, $B_y$, $B_z$) and total ($B_{tot}$) magnetic field; profile of $n$ and $B_{tot}$ as integrated along the transverse direction; ion energy distribution; local Alfv\'en velocity $v_A=B_{tot}/\sqrt{4\pi m_p n}$ in units of the initial one; local inclination of the magnetic field vector with respect to the shock normal (0$^\circ$=parallel, 90$^\circ$=perpendicular).
Such a rich shock structure is entirely generated by accelerated particles, and is dramatically different from the structure of a laminar MHD shock, especially in the upstream \citep{filam}.}
\end{figure}

\begin{figure}
\centering
\includegraphics[width=0.48\textwidth]{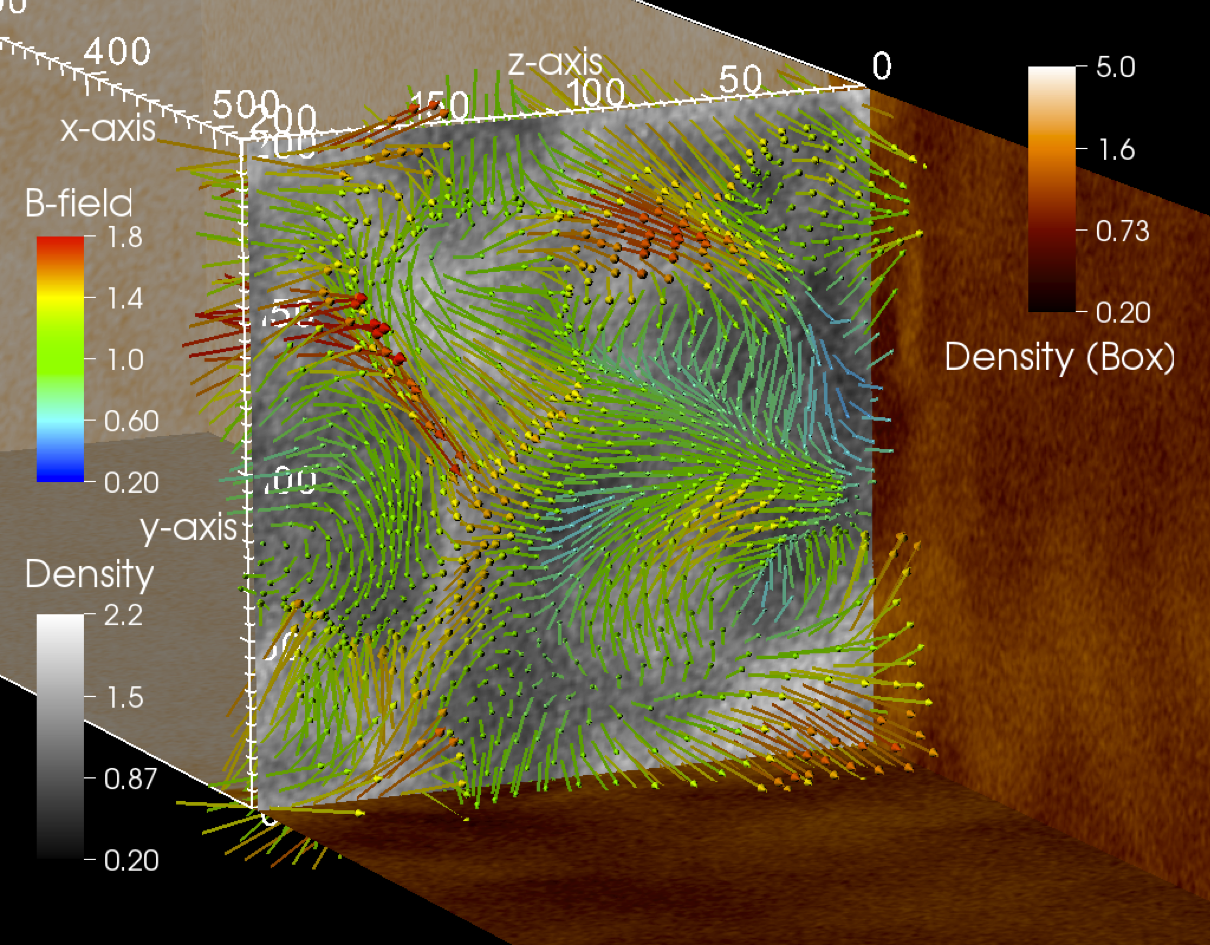}
\caption{\label{fig:3D} \footnotesize
Snapshot at $t=175\omega_c^{-1}$ of a 3D simulation of a parallel $M=6$ shock.
The color code (right colorbar) on the box shows the ion density $n$.
The slice illustrates a section of the fluid about $500c/\omega_p$ ahead of the shock; 
the grey-scale code corresponds to $n$, while the colored vectors show strength and direction of $\vec{B}$, in units of $B_0$ (notice the correlation between $n$ and $B_{tot}$).
The magnetic field is mainly along $\hat{x}$ in the filaments, and coiled around and inside the cavities.
}
\end{figure}

Figure \ref{fig:hybrid} shows the structure of a 2D parallel shock with $M=30$, at $t=500\omega_c^{-1}$  \cite[from][]{DSA}. 
The shock transition is at $x\sim 4000c/\omega_p$, and the upstream (downstream) to the right (left); 
the physical quantities depicted are described in the caption.
Upstream of the shock there is a cloud of high-energy particles (third panel from the bottom in Figure \ref{fig:hybrid}), which drives a current able to amplify the initial magnetic field $\vec{B}=B_0\hat{x}$ by a factor of a few in the shock precursor.
Figure \ref{fig:hybrid} also shows the formation of underdense cavities, surrounded by dense filaments with strong magnetic fields.
The cavities form because the plasma is expelled, along with its frozen-in  field, under the action of a $-\delta \vec{B} \times \vec{J}$ force, where $\vec{J}\parallel \hat{x}$ is the CR current and $\delta \vec{B}$ is the transverse component of the magnetic field, generated via streaming instability.
The net result is that these cavities are filled with energetic particles, which are channeled inside as wires carrying current in the same direction \cite[see also][]{rb13}.
The typical size of the cavities, when they are advected through the shock, is comparable with the gyroradius of the highest-energy particles in the simulation (a few hundred ion skin depths for the simulation shown in Figure \ref{fig:hybrid}).
The 3D topology of the amplified magnetic field in front of the shock is illustrated in Figure \ref{fig:3D} for a $M=6$ parallel shock at time $t=175\omega_c^{-1}$ \cite[from][]{DSA}.

The propagation of the shock through an inhomogeneous medium leads to the formation of turbulent structures in the downstream (Richtmyer--Meshkov instability), in which magnetic fields are stirred, and further amplified.
It is interesting to notice how initial shock strength and topology are dramatically modified by the filamentation instability.
The two bottom panels in Figure \ref{fig:hybrid} show the Alfv\'en velocity, calculated in the local magnetic field $B_{tot}$, and the angle $\vartheta$ between the local field and the $x-$axis. 
The Alfv\'en velocity is typically larger than the initial one, especially around the cavities, making the fluid less super-Alfv\'enic, while the generation of transverse components makes the shock locally oblique over most of its surface (bottom panel of Figure \ref{fig:hybrid}).
As shown in \cite{filam}, cavities always develop at quasi-parallel shocks, along the background magnetic field.
More oblique shocks ($\vartheta\gtrsim 45\deg$), instead, show little or no sign of magnetic field amplification in the shock precursor, due to the lack of accelerated ions diffusing into the upstream \cite{MFA}. 

\subsection{Dependence on Shock Inclination and Strength}
The magnetic fields inferred from X-ray and radio emission in the post-shock regions of young SNRs \cite[e.g.,][]{Tycho} are as large as a few hundred $\mu$G.
In order to obtain such large fields, at the net of the boost of a factor of $\sim 4$ provided by compression at the shock, the interstellar field of a few $\mu$G must be enhanced by factors of several tens \emph{in the shock precursor} \cite[see][for evidence of upstream field amplification]{SN1006}.
Therefore, crucial questions are: how strong can field amplification be at SNR shocks? and what mechanisms are responsible for it?

\begin{figure}\centering
\includegraphics[trim=0px 0px 0px 0px, clip=true, width=.5\textwidth]{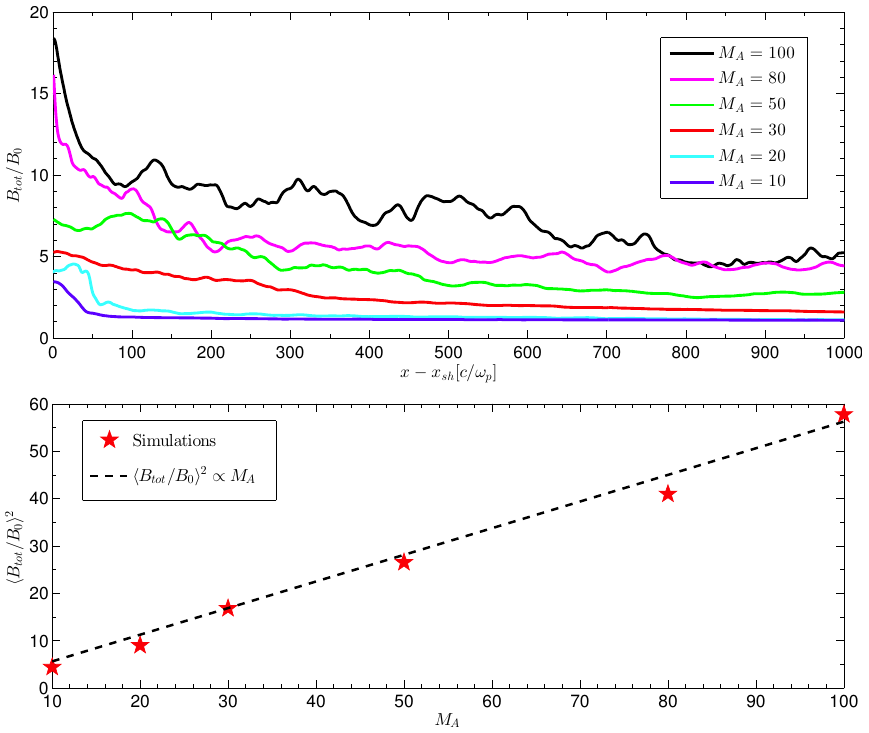}
\caption{\label{fig:dB}\footnotesize
Top panel: upstream profile of the modulus of $\vec{B}$, averaged over $200c/\omega_p$ in the transverse direction, and over $20\omega_c^{-1}$ in time, for different Mach numbers as in the legend; the shock is at $x=0$, and $t=200\omega_c^{-1}$.
Bottom panel: magnetic field immediately ahead of the shock, obtained by averaging the curves in the top panel 
 over a region $\Delta x=10Mc/\omega_p$.
The initial magnetic field $B_0$ is amplified more effectively for large Alfv\'enic Mach numbers, according to $\langle B_{tot}/B_0\rangle ^2\propto M_A$, in good agreement with Eq.~\ref{eq:dB} \citep[from][]{MFA}.}
\end{figure}

In \cite{MFA} we have investigated the dependence of magnetic field amplification on shock strength and inclination;  the main results are summarized in Figure \ref{fig:dB}.
The top panel illustrates the profile of the magnetic field strength in front of the shock, for parallel shocks with Mach numbers up to $M=100$. 
The extent of the region with enhanced field is larger for larger-$M$ shocks, a natural consequence of the fact that accelerated particles have larger energies in units of $mv_A^2$, and in turn larger diffusion lengths in units of ion skin depths $c/\omega_p=v_A/\omega_c$.
Second, and most important, the total amplification factor depends on the shock Mach number. 
The bottom panel of Figure \ref{fig:dB} shows the magnetic energy density at the shock ($\propto B_{tot}^2$), in units of the energy density at upstream infinity, for parallel shocks with $M=10,20,30,50,80,100$ (see the figure caption for more details).
The dashed line passing through the points represents the prediction of the field amplification expected for resonant streaming instability \cite[see, e.g.,][]{bell78a,ab06}, in the following sense.
Introducing the normalized pressure in CRs, $\zeta_{cr}$, defined as the post-shock CR pressure divided by the upstream ram pressure measured in the shock frame, one has:
\begin{equation}\label{eq:dB}
\left\langle \frac{B_{tot}}{B_0}\right\rangle^2\approx 3 \zeta_{cr} \tilde{M},
\end{equation}
where $\tilde{M}\simeq 1.25M$ is Mach number of the upstream fluid in the shock frame \cite[see][for details]{MFA}.
In the range of Mach numbers considered here, $\zeta_{cr}\gtrsim 10\%$ at $t=200\omega_c^{-1}$ (see Figure \ref{fig:eff}); 
the dashed curve in the bottom panel of Figure \ref{fig:dB} corresponds to $\zeta_{cr}=15\%$.
We stress that the fact that magnetic field amplification becomes more prominent for stronger shocks is crucial to account for the large fields inferred in SNRs. 
The typical Mach numbers of young SNRs are as large as a few hundred to thousand: for these shocks Eq.~\ref{eq:dB} returns amplification factors of a few tens, in good agreement with the fields inferred from multi-wavelength observations.

\subsection{Turbulence Spectrum}
In order to characterize the instabilities responsible for magnetic field amplification, we measure the spectrum of the magnetic perturbations in different regions of the shock.
As in \cite{MFA}, we take the Fourier transform of $B_{\perp}(x)$ in the  $k$ space, and express the spectral energy distribution in the magnetic turbulence by introducing $\mathcal{F}(k)$, i.e., the magnetic energy density per unit logarithmic bandwidth of waves with wavenumber $k$, normalized to the initial energy density $B_0^2/(8\pi)$:
\begin{equation}\label{eq:F}
\frac{B_{\perp}^2}{8\pi}=\frac{B_{0}^2}{8\pi}\int_{k_{min}}^{k_{max}}\frac{dk}{k}\mathcal{F}(k).
\end{equation}
We calculate $\mathcal{F}(k)$ in three different regions, which we define as ``far upstream'', ``precursor'', and ``downstream''. 
The upstream is split into a shock precursor, where diffusion in pitch angle is effective and accelerated particles have an almost isotropic distribution function in momentum space, and the far upstream, populated by ions with energy close to the maximum energy $E_{max}$, which escape the system almost free-streaming.
The \emph{free-escape boundary} between the two regions will be better characterized below.

\begin{figure}
\includegraphics[trim=0px 40px 15px 15px, clip=true, width=.49\textwidth]{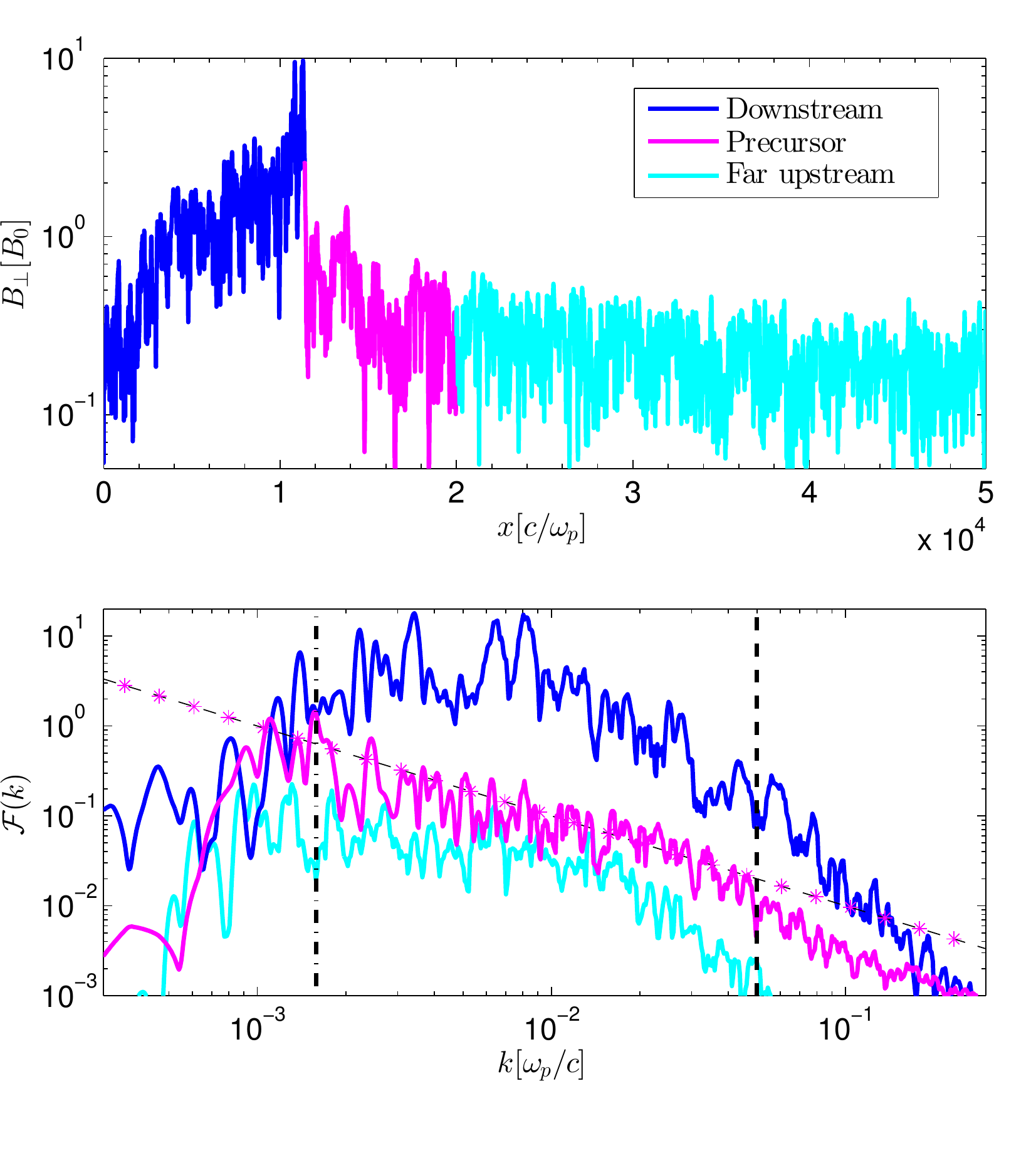}
\caption{\label{fig:Fourier20}\footnotesize
\emph{Top panel}: transverse (self-generated) component of $\vec{B}$ for a $M=20$ parallel shock at $t=2000\omega_c^{-1}$. 
\emph{Bottom panel}: power-spectrum of $B_{\perp}$ as a function of wavenumber $k$.
The color code matches corresponding shock regions.
The vertical dashed and dot-dashed line indicate modes resonant with ions of energy $E_{sh}$ and $E_{max}\sim 300E_{sh}$, respectively.
Symbols correspond to $\mathcal{F}(k)\propto k^{-1}$, i.e., the spectral energy distribution produced by  a $\propto p^{-4}$ CR distribution via resonant streaming instability \cite[from][]{MFA}.}
\includegraphics[trim=0px 40px 15px 15px, clip=true, width=.49\textwidth]{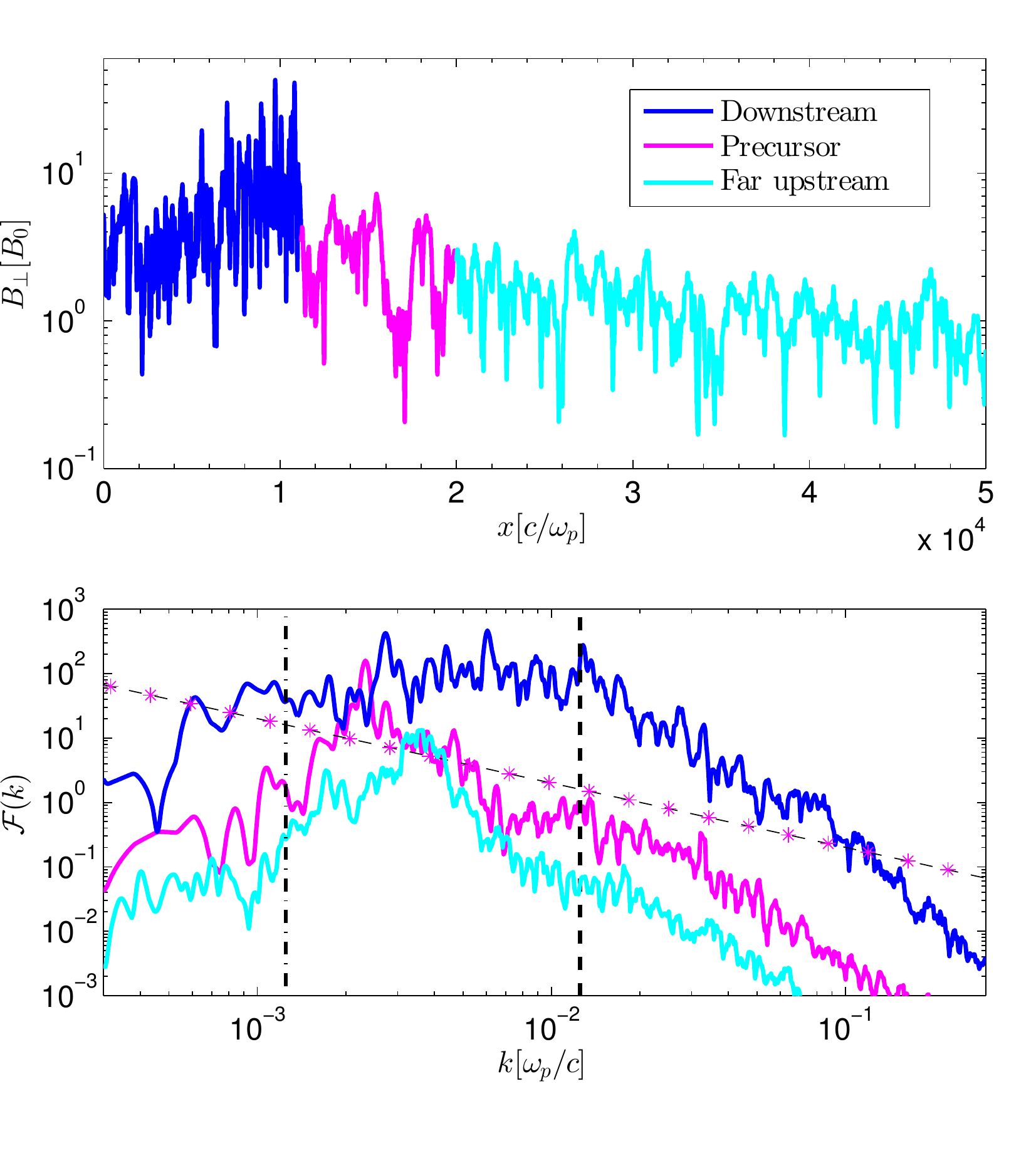}
\caption{\label{fig:Fourier80}\footnotesize
As in figure \ref{fig:Fourier20} for a parallel shock with $M=80$, at $t=500\omega_c^{-1}$, when $E_{max}\approx 100E_{sh}$. 
The magnetic field is significantly more amplified than in the $M=20$ case, with $\mathcal{F}(k)$ in the precursor a factor of almost 10 larger than in  figure \ref{fig:Fourier20}.
Note that the resonance at $E_{max}$ (dot-dashed line) is calculated in $B_0$: accounting for the amplified field would shift the resonance at higher $k$ \cite[from][]{MFA}.}
\end{figure}

The top panel of Figure \ref{fig:Fourier20} shows the space profile of the transverse (self-generated) component of the magnetic field for a parallel shock with $M=20$, and the bottom panel in the same figure shows the magnetic power spectrum; the different curves correspond to the three different regions define above.
The noteworthy points are the following.
In the precursor, the spectrum of the excited modes encompasses the range of wavenumbers $k$ resonant with the momenta of the accelerated particles (between the vertical lines in Figure \ref{fig:Fourier20}), where resonance (in wavelength) is defined by $kr_L(p_k)\approx 1$, with $r_L(p)$ the Larmor radius of ions of momentum $p$ in the background field $B_0$.
In this region, the wave spectrum is $\mathcal F(k)\propto k^{-1}$ (magenta symbols), a trend that matches the energy distribution in the accelerated particles, in the sense that the energy density in waves with wavenumbers in an interval  $dk$ around $k$ is proportional to the energy density in CRs with momentum in a range $dp_k$ around $p_k$.
For a $f(p)\propto p^{-4}$ distribution of non-relativistic ions, the energy density $\propto p$, and the corresponding wave spectrum is expected to go as $k^{-1}$, consistently with our findings\footnote{Note that for the same $p^{-4}$ spectrum of \emph{relativistic} CRs, one would have constant energy density per decade of momentum, and the corresponding wave spectrum would be flat in $k$.}.
Such a ``resonant'' correspondence between energetic ions and the distribution of excited modes is expected in the quasi-linear theory of resonant streaming instability \cite[e.g.,][]{bell78a,ab06}. 
Finally, we notice that the normalization of the wave power spectrum is proportional to the local magnetic field: it is the largest in the downstream, corresponds to $B_{tot}/B_0\gtrsim 1 $ in the precursor, and is smaller far upstream, where the instability had little time for growing.

The situation is quite different for shocks with larger Mach number, where field amplification is more prominent. As an example, in Figure \ref{fig:Fourier80} we show the field profile and the wave spectrum for a parallel shock with $M=80$, where amplification is as large as $B_{tot}/B_0\gtrsim 3$.
As widely discussed in \cite{MFA} (also see \citep{ab09}), for $M\gtrsim30$ the non-resonant hybrid (NRH) instability \cite{bell04} grows significantly faster than the  resonant streaming instability.
At any given time, a fraction of the ions with energy close to $E_{max}$ escape the system because of the lack of waves able to confine them \cite{escape,bell+13}.
In the high-Mach number regime, escaping particles  trigger NRH modes in the far upstream (cyan curves in Figure \ref{fig:Fourier80}); these modes do not effectively scatter ions because the wavelength of the most unstable modes is much smaller than the particles' gyroradius ($k_{max}r_L(E_{max})\gg1$), and have non-resonant polarization \cite[see][for a derivation of their dispersion relations]{ab09}.
The magnetic field quickly grows to non-linear levels, and self-consistent PIC simulations show that when $b\equiv\delta B/B_0\gg 1$, the most unstable mode scales as $k_{max}\propto b^{-2}$, i.e., its wavelength becomes larger and larger \cite{rs09};
at the same time, the ion gyroradius scales as $r_{L}(b)\propto b^{-1}$, so that $k_{max}r_L(E_{max})\propto b^{-3}$. 
For parameters typical of SNR shocks, when $b\gtrsim 5$ the excited modes become resonant in wavelength with $E_{max}$ ions, which are thus effectively scattered in pitch angle \cite[see][]{MFA,bell+13}.
The location in the upstream where this confinement is realized corresponds to the free-escape boundary mentioned above, and marks the separation between the far upstream and the precursor.
Between the free-escape boundary and the shock, magnetic field amplification is provided by the current in diffusing CRs, as in standard CR-dominated shock precursors \cite[e.g.,][]{bell78a,ab06,efficiency}.
The peak in the wave power spectrum (bottom panel of Figure \ref{fig:Fourier80}) shifts to lower $k$  when moving from far upstream toward the shock.
When NRH modes are prominent, the wave spectrum at the shock appears quite different from the $\mathcal{F}(k)\propto k^{-1}$ prediction of resonant streaming instability, being the result of the convolution of modes excited in different upstream regions, where turbulence generation is strongly non-linear.

\section{\label{sec:diff}Particle Diffusion}
In addition to being necessary for explaining the synchrotron emission from young SNRs, magnetic field amplification is also required for enhancing CR scattering, and in turn favoring their acceleration.
CRs are scattered in pitch angle by collisions against waves with resonant wavelengths, and this process can usually be described by introducing a diffusion coefficient. 
A popular choice is to assume that the mean free path is comparable with the particle's gyroradius (\emph{Bohm diffusion}), with a diffusion coefficient that reads:
\begin{equation}\label{eq:DB}
D_B(E)\equiv\frac{v}{2}r_L(v,B)=\frac{v}{2}\frac{pc}{eB}=\frac{E}{m\omega_c}\,,
\end{equation}
where $v$ is the particle velocity\footnote{In the denominator there is usually a factor 3: here we pose 2 because we consider 2D simulations \cite[see][for more details]{diffusion}}.
From the kinetic point of view, Bohm diffusion is achieved in the turbulence generated by a $f(p)\propto p^{-4}$ particle distribution via resonant streaming instability, in the quasi-linear limit $\delta B/B_0\sim 1$ \cite[see, e.g.,][]{bell78a}.
This diffusion coefficient is often heuristically extrapolated into the regime of strong field amplification, scaled as $D_B(b)=D_B(B_0)/b$, but such a prescription lacks a solid theoretical justification.  

\begin{figure}\centering
\includegraphics[trim=0px 30px 0px 260px, clip=true, width=.485\textwidth]{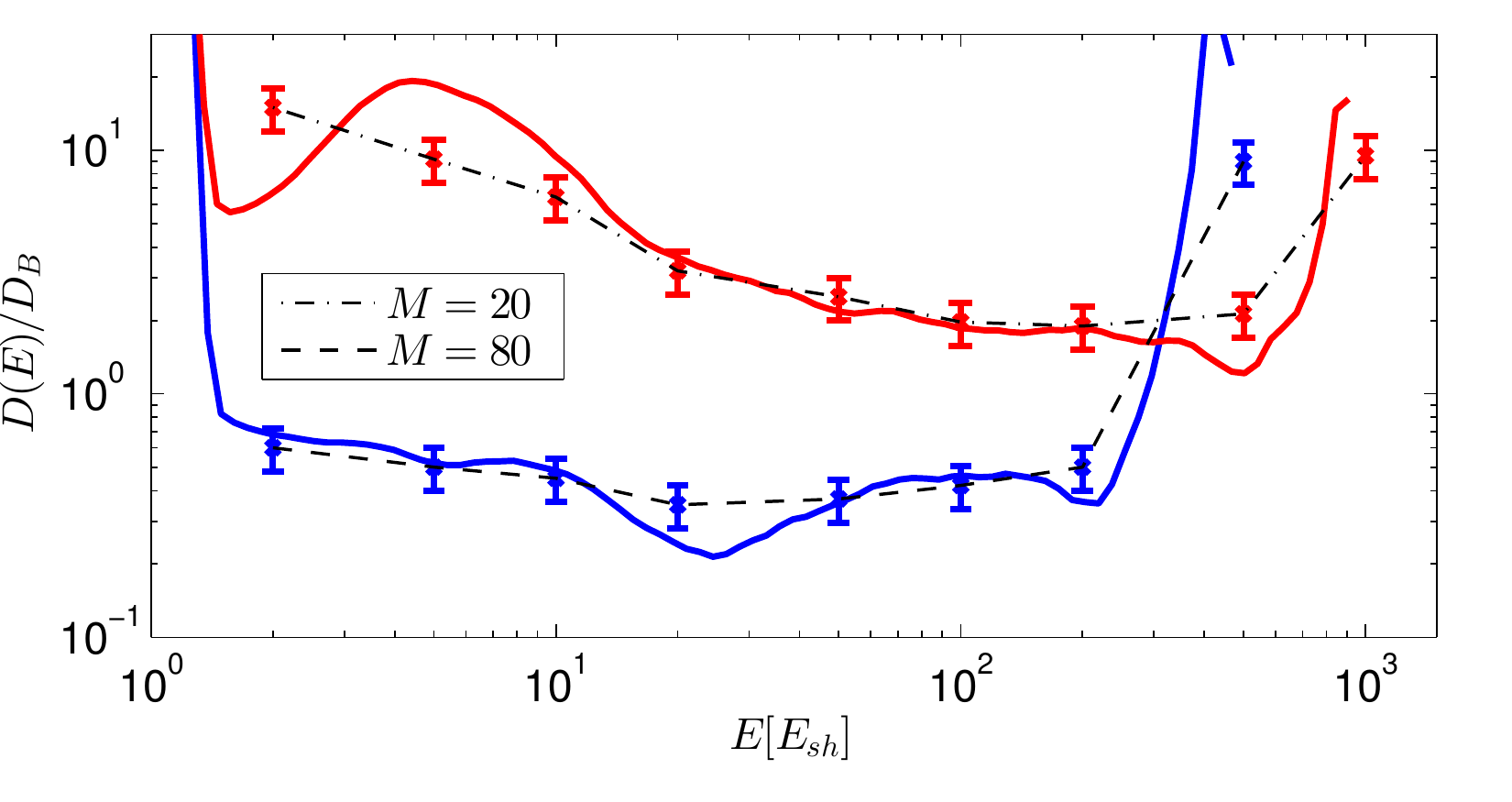}
\caption{\label{fig:cd2080}\footnotesize
Diffusion coefficient, normalized to Bohm, immediately in front of the shock for $M=20,80$, inferred by tracking individual particles (points with fiducial error bars of 20\%), and by using the analytical procedure outlined in \cite{diffusion} (solid red and blue lines).}
\end{figure}
\begin{figure}
\includegraphics[trim=0px 280px 0px 15px, clip=true, width=.485\textwidth]{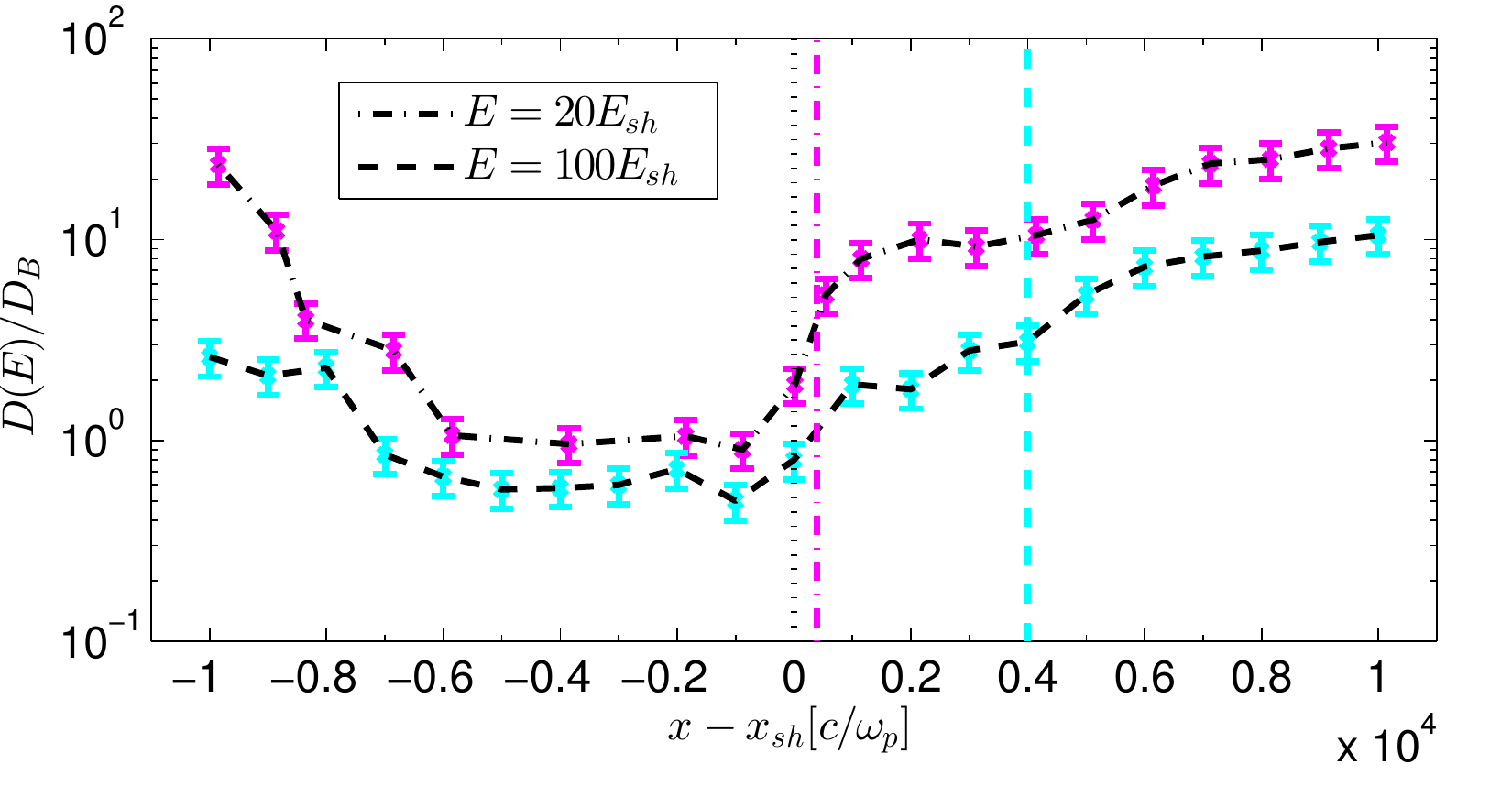}
\caption{\label{fig:cdx}\footnotesize
Spatial dependence of the diffusion coefficient for a shock with $M=20$ at $t=2400\omega_c^{-1}$, for particles with energy $E=20,100E_{sh}$ as in the legend. $D(E,x)$ is calculated by tracking CRs in periodic boxes centered at $x$, and of width indicated as the distance between the vertical colored lines and the dotted line \cite[see][for more details]{diffusion}.}
\end{figure}

Our global, self-consistent simulations allow to reconstruct the properties of particle diffusion in different regions of the shock, and to compare them with Bohm diffusion, or with the more refined prediction of the diffusion rate in the presence of Alfv\'enic modes with spectrum $\mathcal{F}(k)$ \cite{bell78a,ab06}.
In the case of waves excited by accelerated particles themselves, the so-called \emph{self-generated} diffusion coefficient reads \citep[see, e.g.,][]{bell78a}:
\begin{equation}\label{eq:Dsg}
D_{sg}(p)=\frac{8}{3\pi}\frac{D_B(p)}{\mathcal{F}(k_p)},
\end{equation}
where $k_p$ is the resonant wavenumber, as defined above. 
From Equation \ref{eq:Dsg} one sees that, since $\mathcal{F}(k)\propto (\delta B(k)/B_0)^2$, strong magnetic turbulence suppresses the diffusion coefficient at the corresponding scales (i.e., resonant in momentum).
We measured the local diffusion coefficient in two different ways, which are extensively discussed in \cite{diffusion}: 
i) by using an analytical procedure based on the extent of the CR distribution in the upstream, at any given momentum; 
ii) by tracking individual test particles in boxes initialized with the turbulence pertaining to different shock regions.
The two methods return consistent diffusion coefficients, which are shown in Figure \ref{fig:cd2080} as a function of ion energy, and for two shocks with $M=20$ and $M=80$ \cite[see][for details]{diffusion}.
Two things have to be noticed here. 
First, the energy dependence of the diffusion coefficient is quite different for $M=20$ and $M=80$.
For moderately strong shocks, where magnetic field amplification occurs in the quasi-linear regime, one finds $D(p)\propto p$ (red line in Figure \ref{fig:cd2080}), as a consequence of having $\mathcal{F}(k)\propto k^{-1}\propto p$ in Eq.~\ref{eq:Dsg}.
For stronger shocks, like for $M=80$, the inferred diffusion coefficient is roughly proportional to the Bohm coefficient; indeed, field amplification is strongly nonlinear, and Eq.~\ref{eq:Dsg} is not expected to hold.
In both cases, the diffusion coefficient increases abruptly above the maximum energy in the CR distribution, because of the lack of wave generation at the corresponding resonant scales.
Second, the overall normalization depends on the level of magnetic field amplification, and for $M=80$ is smaller than for $M=20$, approximately by the ratio of the field amplification factor, which is a factor of a few (see also Figure \ref{fig:dB}).

We have also investigated via particle tracking the spatial dependence of the diffusion coefficient for a parallel shock with $M=20$. 
Figure \ref{fig:cdx} shows the diffusion coefficient for ions with $E_{sh}=20,100$ in different shock regions: diffusion is enhanced where the magnetic field is stronger, namely in the downstream, and in the shock precursor because of self-generated fields.
We conclude that, while for moderately-strong shocks the quasi-linear theory of ion diffusion in the self-generated field does apply, at strong shocks with $M\gtrsim 30$ ion diffusion is instead well-described by Bohm diffusion, calculated in the amplified magnetic field.

\section{\label{sec:Emax}Maximum Energy}
\begin{figure}\centering
\includegraphics[trim=30px 0px 40px 5px, clip=true, width=.485\textwidth]{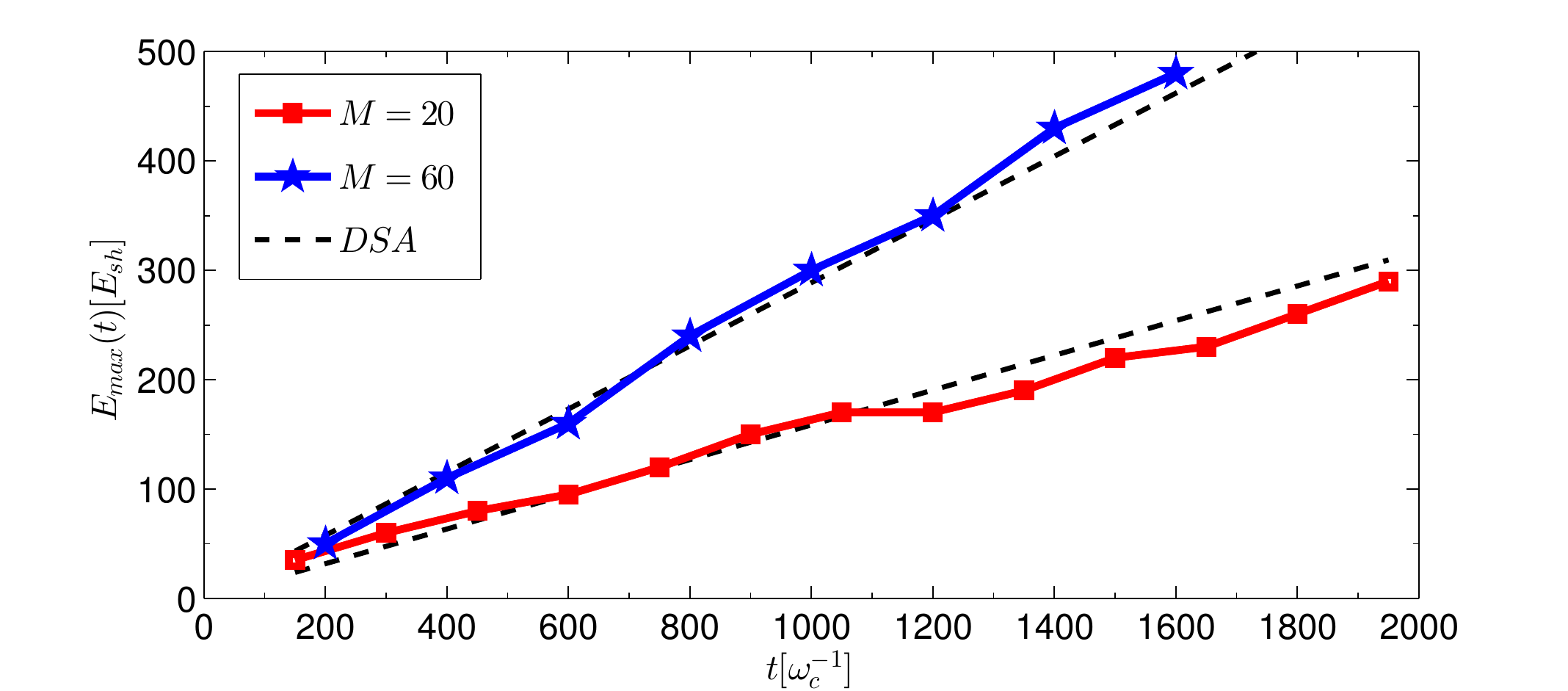}
\caption{\label{fig:Emax}\footnotesize
Time evolution of the maximum ion energy for parallel shocks with $M=20$ and 60, compared with the DSA prediction  according to Eq.~\ref{eq:Emax_B}, with $\kappa_{20}= 2.1$ and $\kappa_{60}=$1.2, respectively (dashed lines) \cite[see][for more details]{diffusion}.}
\end{figure}

Enhanced scattering favors ion return to the shock, and actually determines the maximum energy that can be achieved in a given time.
For DSA, the acceleration time is promptly written as a function of the diffusion coefficient \cite[e.g.,][]{drury83,bac07}, and eventually the expected time scaling of the maximum energy $E_{max}(t)$ reads:
\begin{equation}\label{eq:Emax_B}
E_{max}(t)\simeq \frac{E_{sh}}{3\kappa}\omega_c t,
\end{equation}
where we introduced $\kappa\equiv D(E_{max})/D_B(E_{max})$ to express the deviation with respect to the DSA prediction with Bohm scattering.
Figure \ref{fig:Emax} shows the evolution of $E_{max}$, found by fitting the post-shock ion spectrum with a power-law plus an exponential cut-off, for two parallel shocks with $M=20,60$, as discussed in \cite{diffusion}. 
We leave $\kappa$ as a free parameter, and the best-fitting curves passing through the points in Figure \ref{fig:Emax} correspond to $\kappa_{20}\sim 2.1$ for $M=20$, and  $\kappa_{60}\sim 1.2$ for $M=60$.
These values provide another, independent, estimate of the value of the diffusion coefficient close to $E_{max}$, and only differ by a factor of about 2 from the instantaneous values of $D(E)$ illustrated above. 
Such a discrepancy may be due to the fact that all the relevant quantities, such as $B_{tot},\mathcal{F},D(E_{max})$, are actually function of time and position.
In any case, our findings attest to the decrease of the acceleration time with the increase of magnetic field amplification, suggesting that strong shocks with large $\delta B/B_0$ can accelerate ions to energies much larger than those achievable with Bohm diffusion in the unperturbed magnetic field.

\section{Conclusions}
We performed an extended investigation of the fundamental mechanisms responsible for ion acceleration and magnetic field generation in non-relativistic collisionless shocks, by means of unprecedentedly-large hybrid simulations.
Since ions are treated kinetically from first principles, these simulations return global electromagnetic shock structures, in which ion injection, acceleration, and escape are treated self-consistently.
We find that at quasi-parallel strong shocks ions are accelerated via DSA, with efficiencies as large as 10--20\%; the spectrum of accelerated ions agrees with the  DSA universal prediction of $f(p)\propto p^{-4}$ (\S\ref{sec:acc}).
These values are close to the ones inferred via $\gamma$-ray observations of young SNRs, even if ion spectra are often slightly steeper (with typical spectral indexes between ${4.2}$ and ${4.5}$), which suggests that non-linear corrections to standard DSA may be required \cite[see][for an extended discussion]{efficiency}.  

We attested to the relevance of filamentation, resonant, and non-resonant hybrid instabilities in amplifying the initial magnetic field, and showed that the total amplification factor scales with the square root of the Alfv\'enic Mach number up to $M=100$.  
We characterized the extent of the shock precursor, i.e., the region where energetic ions diffuse, determining the position of the free-escape boundary, from which highest-energy ions leave the system because of lack of confinement (\S\ref{sec:MFA}).
Finally, we showed that particle diffusion occurs close to the Bohm limit, i.e., the mean free path for pitch angle scattering is comparable with the ion gyroradius in the amplified field (\S\ref{sec:diff}). Such enhanced scattering favors the fast energization of accelerating ions, and determines the evolution of the ion maximum energy (\S\ref{sec:Emax}).

\subsection*{Acknowledgments}
\footnotesize
We thank L.\ Gargat\'e for providing a version of \emph{dHybrid}.
This research was supported by NSF grant AST-0807381 and NASA grant NNX12AD01G, and facilitated by the Max-Planck/Princeton Center for Plasma Physics.
Simulations were performed on the computational resources supported by the PICSciE-OIT TIGRESS High Performance Computing Center and Visualization Laboratory. This research also used the resources of the National Energy Research Scientific Computing Center, which is supported by the Office of Science of the U.S. Department of Energy under Contract No.\ DE-AC02-05CH11231, and XSEDE's Stampede under allocation No.\ TG-AST100035.
\nocite{*}
\bibliographystyle{elsarticle-num.bst}
\bibliography{bibsanvito}







\end{document}